# The quantitative side of the Repertory Grid Technique: some concerns

**Evangelos Karapanos**

User-Centered Engineering

Department of Industrial Design

Eindhoven University of Technology

P.O. Box 513, 5600 MB, Eindhoven,

The Netherlands

E.Karapanos@tue.nl

**Jean-Bernard Martens**

User-Centered Engineering

Department of Industrial Design

Eindhoven University of Technology

P.O. Box 513, 5600 MB, Eindhoven

The Netherlands

J.B.O.S.Martens@tue.nl

**Abstract**

User experience (UX) evaluation is gaining increased interest lately, both from academia and industry. In this paper we argue that UX evaluation needs to fulfill two important requirements: *scalability*, i.e. the ability to provide useful feedback in different stages of the design, and *diversity*, i.e. the ability to reflect the diversity of opinions that may exist in different users. We promote the use of the Repertory Grid Technique as a promising UX evaluation technique and discuss some of our concerns regarding the quantitative side of its use.

**ACM Classification Keywords**

H5.2. User Interfaces: Evaluation/methodology.

**Introduction**

User experience (UX) evaluation has become a popular topic in recent research and is gaining increased interest from industry. UX aspects of products are not only important at the time of product purchase, but continue to influence customer satisfaction throughout a product's lifecycle [5]. In a recent study that was conducted in collaboration with a multi-national consumer electronics company, it was found that UX aspects are be-coming increasingly important. For example, it was estimated that almost 50% of product returns in 2002 could be attributed to products not meeting UX expectations [17]. Reasons for such user dissatisfactions were attributed to a mismatch between product specifications (i.e. what the designers intended to deliver) and the actual user needs and desires. In a root-cause analysis it was found that decisions made in the conceptual design phase had the highest impact on the emergence of UX failures. This means that it is important to not only measure UX aspects of finished products, but also early on in the design process, i.e. using early prototypes or scenarios. This in turn implies that UX measurement methods will need to be *scalable,* i.e., suited to provide useful feedback in all stages of the design process. Successive generations of a product under development allow for different insights, for instance, the interaction with a product may not be evident in the concept evaluation phase, since it is typically designed only in a later stage. It is thus important for UX evaluation to not only be *iterative* but also *feedforward,* in the sense that it needs to provide insight about aspects yet to be designed (e.g. what is the product's character and how should the interaction be designed to comply with it).

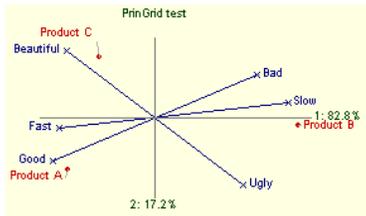

**Figure 1.** An individual's scores for products A to C based on the three personal attributes. Scores range from 1 to 5.

**Figure 2.** A Principal Components Analysis on the individual's judgments.

Another requirement for UX evaluation is the ability to cope with diversity. Cooper [1] noticed that not all users need to have a positive experience for a product to be successful. It is thus important for UX evaluation to understand the diverse experiences so that the product can be fine-tuned for an optimal experience for a specific audience.

We intend to show that the Repertory Grid Technique (RGT) satisfies many of the above-mentioned requirements. In our experience using the RGT we have however also developed some concerns about the more quantitative aspects of the technique. In the rest of the paper we introduce the technique and discuss some of these concerns.

## The Repertory Grid Technique

The Repertory Grid Technique (RGT) [4] is a structured interview approach, in which insight is gained into the ways that people distinguish between different elements of choice, by comparing them in sets of three, the so called triads. RGT consists of two phases: (a) elicitation of user perceived attributes (the so-called personal constructs) and (b) rating on the elicited attributes.

In the first phase, three different elements are presented to the participant and (s)he is asked to "think of a property or quality that makes two of the elements alike and discriminates them from the third". The result is a set of (often bi-polar) attributes such as "easy to use – hard to use". The elements presented can vary from usage scenarios to vertical prototypes. At least four elements are required for the creation of more than one triad.

In the second phase, the participant is asked to rate all elements on his/her own elicited attributes. The bi-polar attributes are often employed in seven or nine point semantic differential scales [11] and the participant is rating each element individually on the resulting scales. The ratings can be summarized in a matrix (Fig. 1), where the columns correspond to different elements and the rows correspond to different attribute ratings. Relationships between attributes and elements can for instance be explored by means of (hierarchical) clustering, factor analysis or multidimensional scaling (Fig. 2).

## RGT as an evaluative tool

The RGT can serve different purposes in UX research and practice. It can be applied as an *inspirational* tool, in which case the emphasis is on rich insight, and the irrationalities in elicited attributes (e.g. beautiful – square shape) are respected, since they might provide insight into the idiosyncratic ways in which users infer overall quality from concrete design attributes. On the other hand, the RGT can also be applied as an *evaluative* tool, in which case the emphasis is on acquiring a valid and reliable representation of the user experience. In our experience using the RGT as an evaluative tool we have felt the need to modify parts of the technique. We have come to see that certain principles, which are dragged along with the adoption of the tool, are not necessarily valid in the field of user experience. In the next paragraphs we present some of the modifications we propose regarding both the quantitative and the qualitative side of the RGT analysis.

## Are we really interested in idiosyncratic views?

The RGT originates from clinical psychology where the emphasis is on an individual's perceptual and cognitive processes. In the field of user experience, however, the interest is not in the idiosyncratic views of an individual but rather on some more-or-less homogeneous groups of individuals. Due to this focus on idiosyncrasy, interpersonal analysis of the RGT has received very little attention. We have come to realize two inadequacies of the RGT when used for interpersonal analysis.

First, the interpersonal classification of attributes is often performed purely on semantic grounds, i.e., without testing congruence on the rating scores. Such a practice does not take properly into account the diverse ways in which individuals construe attributes to refer to internal concepts [4].

Second, techniques such as Principal Components Analysis (PCA) or Multi-Dimensional Scaling (MDS) assume homogeneity in the way people perceive the elements offered to them. This is sometimes referred to as the *principle of homogeneity of perception* [10]. Due to the lack of methodological support for the interpersonal analysis of RGT data, averaging across individuals becomes a common practice. This is not in agreement with the original motivation of the RGT, which is to account for diversity in individuals' judgments. A modified MDS data analysis technique that can account for such diversity in RGT data has therefore recently been proposed [6, 8].

## On bipolarity

A central notion in the RGT is the bipolarity of the idiosyncratic constructs (i.e. attributes) Kelly, in his theory of Personal Constructs, postulated that individuals perceive the world around them through the construction of dichotomous constructs. It is our experience, however, that subjects often need to be probed in order to derive a truly bipolar attribute. This raises concerns with respect to whether individuals actually do think in bipolar terms. Lyons [9] posited that *"categorizing experience in dichotomous contrasts is a universal human tendency which is only secondarily reflected in language"*. He identified three different types of bipolarity: *negation* (i.e. practical-impractical), *opposition* (i.e. professional - amateurish) and *non-contiguous*, where the opposite pole does not constitute a negation or linguistic opposition (i.e. easy – powerful) (cf. [12]).

**Table 1.** Percentages of attributes types from Karapanos & Martens [7] and Hassenzahl & Wessler [4] studies.

| Type | [7] | [4] |
|---|---|---|
| Negation | 67% | 35% |
| Opposition | 17% | 26% |
| Non-contiguous | 16% | 39% |

**Table 1.** Amount of variance accounted for by the latent one-dimensional construct. Original attributes were in Dutch.

| Attribute | $R^2$ |
|---|---|
| *Negation* | |
| Secure | 0.88 |
| Practical | 0.86 |
| *Opposition* | |
| Creative | 0.75 |
| Professional | 0.90 |

In a meta-analysis of a RGT study on early concept evaluation [7], we found that the majority of the elicited constructs (67%) were negation constructs, while 17% were opposition constructs and only 16% were non-contiguous. This deviates from what was observed in a study by Hassenzahl and Wessler [4] where non-contiguous constructs constituted the major category (39%) while negation accounted for 35% of the constructs and opposition for the remaining 26% (these percentages were for two products in the reported study). This observed discrepancy is likely to have been influenced by two aspects: a) the instructions of the experimenter (our study had a more evaluative character, while the latter study focused on informing design) and b) the fidelity of the prototypes (in our study early concepts were communicated in sketched scenarios, while in the latter study users interacted with working prototypes).

It becomes apparent that the application of the RGT needs to comply with its purpose, *inspirational* or *evaluative*. If the scope is to inspire design, emphasis should be placed on eliciting non-contiguous constructs that provide insight into the relationships that individuals perceive between design qualities (i.e. beautiful – hard-to-use) and concrete product attributes (i.e. many buttons). If the RGT has an evaluative tone, however, non-contiguous constructs create complications. In our experience we have seen that it is often the case that, during rating, subjects cannot recall the context in which the attributes were elicited. Furthermore, attributes elicited within a certain triad can not necessarily be applied to all products. When both poles of a bipolar construct are not (equally) evident to the subject, ratings may very well be based mostly on one of the poles. We would thus suggest that attributes should be validated by the subject before moving to the rating phase. In this attribute validation phase, subjects can be asked to remove duplicate attributes and rephrase attributes when needed.

In the rating process, non-contiguous attributes will evidently elicit distorted ratings since the two poles do not underlie a single one-dimensional construct. Negation and opposition bipolarity constitute the common practice in validated questionnaires. In a small study we attempted to explore the difference, if any, between negation and opposition bipolarity in rating scales. Our interest was to test the opposition hypothesis, i.e. ratings for the negative and positive pole should have a linear correlation of -1. Fourteen subjects rated three concepts on two negation (i.e. secure-unsecure, practical-impractical) and two opposition (i.e. standard-creative, amateurish-professional) attributes, using 7-point semantic differential scales. All attributes except the one referring to security were selected from Attraciff2 [3], a validated user experience questionnaire. These three attributes were also identified in a RGT study with the three concepts, ensuring the content validity of the scales for the current study.

The opposition hypothesis was tested by attempting two different Multi-dimensional Scaling models, a one-dimensional and a two-dimensional. The two-dimensional model provided better fit for all four attributes ($\chi^2$, $p<0.001$), implying that there is significant evidence for the fact that the ratings for the positive and negative poles of the attributes do not underlie a single latent one-dimensional construct. Table 2 depicts the squared multiple correlation coefficient $R^2$ between the latent one-dimensional construct and the original attribute. One can note that no clear differences be-tween negation and opposition attributes emerged in this limited study. The attribute *standard-creative* (original term in Dutch fantasieloos-creatief) shows the least agreement between the two opposite poles.

**On the measurement of meaning**
Due to the notion of bipolarity, semantic differential [11] scales are traditionally employed in RGT rating practices. We have been promoting comparison scales (such as paired comparisons) [6] as an alternative, since such scales are known to be less sensitive to contextual effects than single-stimulus scales, such as the semantic differentials. De Ridder [2] explored this issue in an experiment where quality assessments were made on a set of images. The same set of images was embedded in a larger set of images with a positively or a negatively skewed distribution in quality. The sensitivity of different scales (such as single-stimulus and comparison scales) to these contextual effects was assessed. It was found that contextual effects were negligible only when comparison scales were used. Thus, in the context of the RGT, contextual effects such as individual differences in prior experiences or order effects during the rating process will evidently make individuals' judgments less reliable when single-stimulus scales are employed.

**Conclusion**
The RGT for assessing user experience can serve two complementary roles: *inspirational* and *evaluative*. When the RGT is used as an evaluative technique one should refrain from averaging across users, question the bipolarity of attributes, and avoid the use of single-stimulus scales. Some of the proposed modifications to the RGT should help to make it a more useful and reliable technique for measuring user experiences.

**References**
[1] Cooper, A., The Inmates Are Running the Asylum, Macmillan Publishing Co., Inc., 1999.

[2] de Ridder, H., Current issues and new techniques in visual quality assessment, Image Processing, 1996. Proceedings., International Conference on, 1 (1996).

[3] Hassenzahl, M., The interplay of beauty, goodness, and usability in interactive products, Human-Computer Interaction, 19 (2004), pp. 319-349.

[4] Hassenzahl, M. and Wessler, R., Capturing design space from a user perspective: The Repertory Grid Technique revisited, International Journal of Human-Computer Interaction, 12 (2000), pp. 441-459.

[5] Hummels, C. C. M., Searching for salient aspects of resonant interaction, Knowledge, Technology and Policy, 20 (2007).

[6] Karapanos, E. and Martens, J.-B., Characterizing the diversity in users' perceptions, in C. Baranauskas, ed., *Human-Computer Interaction - INTERACT 2007*, Springer, 2007, pp. 515-518.

[7] Karapanos, E. and Martens, J.-B., On the discrepancies between designers' and users' perceptions as antecedents of failures in motivating use, in K. Blashki, ed., *International Conference Interfaces and Human Computer Interaction*, IADIS, Lisbon, 2007, pp. 206-210.

[8] Karapanos, E., Martens, J.-B. and Hassenzahl, M., Accounting for diversity in subjective judgments (submitted), *CHI'08*, 2008.

[9] Lyons, J., Semantics, Cambridge University Press, 1977.

[10] Martens, J.-B., Image technology design: A perceptual approach, Kluwer Academic Publisher, Boston, 2003.

[11] Osgood, C. E., Suci, G. and Tannenbaum, P., The measurement of meaning, University of Illinois Press, Urbana, IL, 1957.

[12] Yorke, M., Bipolarity or not? Some Conceptual Problems Relating to Bipolar Rating Scales, British Educational Research Journal, 27 (2001), pp. 171-186.